\documentclass{iaus}
\usepackage{graphicx}
\usepackage{natbib}
\bibpunct{(}{)}{;}{a}{}{,}
\def\lsol  {\hbox{L$_{\odot}$}}
\def\irlum {\hbox{$L_{\rm FIR}$}}
\def\HCOP  {\hbox{HCO$^+$}}
\def\water {\hbox{${\rm H}_2{\rm O}$}}

\title[Molecular properties of (U)LIRGs] 
{Molecular properties of (U)LIRGs:\\ CO, HCN, HNC and \HCOP}

\author[Loenen, Baan \& Spaans]   
{A.F. Loenen$^{1,2}$, W.A. Baan$^2$ \and M. Spaans$^1$}

\affiliation{$^1$Kapteyn Astronomical Institute, University of Groningen,
\break P.O. Box 800, 9700 AV Groningen, The Netherlands\break email:
loenen@astro.rug.nl, spaans@astro.rug.nl\\[\affilskip]
$^2$ASTRON,  P.O. Box 2, 7990 AA Dwingeloo, The Netherlands \break
email: baan@astron.nl}

\pubyear{2007}
\volume{242}  
\pagerange{??--??}
\date{?? and in revised form ??}
 \jname{Proceedings Astrophysical Masers and
their Environments} \editors{J.M. Chapman \& W.A. Baan, eds.}
\begin{document}

\maketitle

\begin{abstract}
The observed molecular properties of a sample of FIR-luminous and OH
megamaser (OH-MM) galaxies have been investigated. The ratio of high
and low-density tracer lines is found to be determined by the
progression of the star formation in the system.  The \HCOP/HCN and
\HCOP/HNC line ratios are good proxies for the density of the gas, and
PDR and XDR sources can be distinguished using the HNC/HCN line
ratio. The properties of the OH-MM sources in the sample can be
explained by PDR chemistry in gas with densities higher than
$10^{5.5}$ cm$^{-3}$, confirming the classical OH-MM model of IR
pumped amplification with (variable) low gains.

\end{abstract}

\firstsection 
\section{Introduction}
\label{sec:introduction}
High-density molecular gas plays an important role in the physics of
(Ultra-)Luminous Infrared Galaxies ((U)LIRGs), giving rise to
spectacular starbursts and possibly providing the fuel for an active
galactic nucleus (AGN).  The emission lines emanating from the nuclear
gas provide information about the physical properties of the nuclear
environment in these systems, e.g. the (column) density, temperature
and chemical composition of the gas, and the type and strength of the
central energy source. It can also provide us an insight into the
processes influencing the (gas in the) nuclei: the star formation rate
and history, fuelling of a possible central black hole and feedback
processes.

\cite{baanHLBW07} present data of the CO, HCN, HNC, \HCOP, CN, and CS
line emissions of a representative group of 37 FIR-luminous and OH
megamaser (OH-MM) galaxies and 80 additional sources taken from the
literature. In this work, the molecular characteristics of this sample
are explained using several models.  First, the properties of the
different (density) components of the nuclear gas are explained in
terms of starburst evolution \citep[see][]{baanHLBW07}. Also the
chemical properties of the high-density gas are analyzed, using
chemical networks \citep{meijSpaans05} and radiative transfer models
\cite{meijSI07}. Here only the J=1-0 transitions of the molecules are
considered. A more detailed analysis will be presented in
\cite{loenenBS07}.



\section{Starburst evolution}
\label{sec:starburst-evolution}
\begin{figure}
\begin{minipage}[t]{0.64\textwidth}
\includegraphics[width=.485\textwidth]{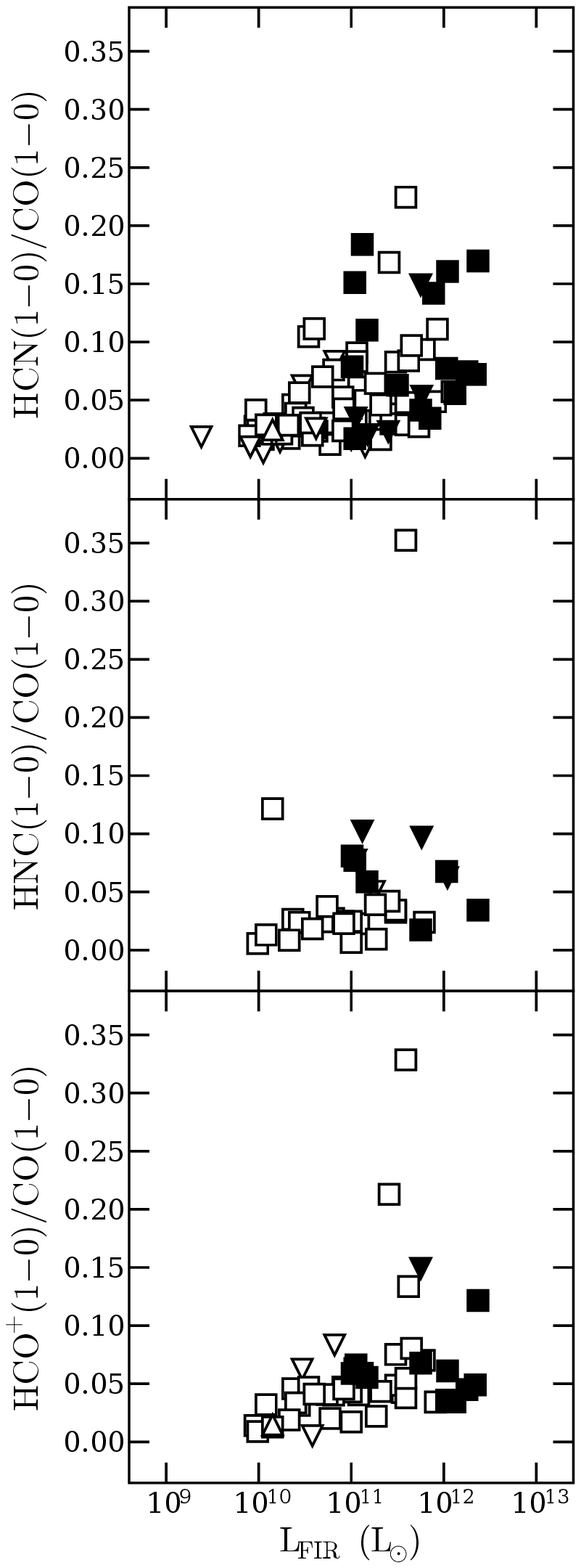}
\includegraphics[width=.515\textwidth]{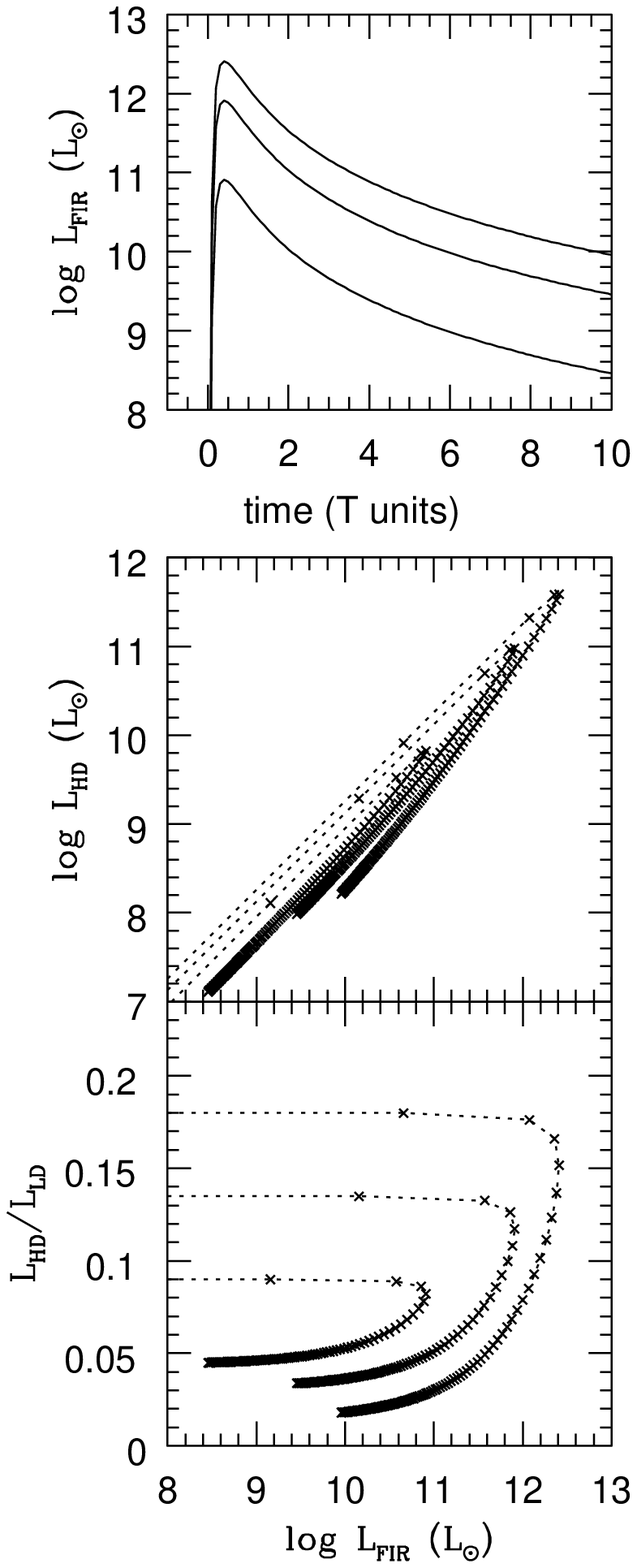}
\end{minipage}\hfill
    \begin{minipage}[t]{0.3\textwidth}
      \vspace{-10.5cm}
      \caption{\textit{left, from top to bottom:} Integrated line
ratios HCN(1-0)/CO(1-0), HNC(1-0)/CO(1-0), and \HCOP(1-0)/CO(1-0)
versus FIR luminosity. Squares represent reliable values and triangles
upper or lower limits. Filled symbols are sources with OH-MM activity.
\newline\textit{right, from top to bottom:} Three FIR luminosity
curves used in the simulations with different maximum luminosities,
the luminosity of a high-density component, and the variation of the
high- versus low-density ratio during the outburst starting at the
upper part of the curve. \label{fig:SB-evo}}
    \end{minipage}
\end{figure}

%
The gas in galaxies is build up from multiple components, each with
different densities and temperatures.  From our data we can derive
information about these different components.  The CO(1-0) line traces
the large scale low-density (critical density $n_{\rm
cr}$=$3\times10^3$ cm$^{-3}$) component (LD), whereas the lines of
HCN, HNC, and \HCOP\ (all $n_{\rm cr}$$>$$10^5$ cm$^{-3}$) trace the
high-density component (HD) which mostly resides in the cores of the
galaxies, at the sites of the star-formation activity.

On the left hand side of Fig.~\ref{fig:SB-evo}, the relative
contributions of these two density components are shown, by plotting
the ratios of the high-density and CO(1-0) line strengths. The figure
shows that the distribution of line ratios for all molecules increases
with FIR luminosity, which gives an upwardly curved lower boundary for
the distribution at higher \irlum. The highest values are found at
\irlum$\ge$$10^{10.5}$ \lsol. The figure also shows that in general the
OH-MM sources have a much larger spread in HD/CO(1-0).

This behavior can be explained as the result of ongoing star
formation.  The FIR luminosity of the ULIRG during the evolution of
the outburst reflects energy generated by the star formation
activity. The FIR luminosity integrated over the course of the
outburst would reflect the amount of high-density molecular material
consumed by the star formation process and destroyed or removed by
feedback.

In the following simplified scenario \citep[see][]{baanHLBW07}, we
employ a model for the time-evolution of the high-density components
in a galaxy during a FIR outburst.  In the absence of a representative
FIR light curve, we use a diffusion-like expression in time $t$ as a
response to a starburst starting at $t$=0 defined as:
\begin{equation}
L_{\rm FIR}(t) = 1.35 L_{\rm FIR}(0) \left(\frac{t}{T}\right)^{2.5}
e^{-t/T}~,
\label{eq:1}
\end{equation}
where $L_{FIR}(0)$ is the maximum luminosity of the burst and $T$ is
the timescale of the outburst.  We note that a diffusion curve may not
be the most appropriate representation of $L_{\rm FIR}$, but this
curve does resemble the outcome of starburst-driven FIR evolution
simulations  \citep{loenenBS06}.

The high-density component HD can be defined as $\beta$ LD, the
low-density component. As a result, the high-low-density ratio varies
with time during the FIR outburst as:
\begin{equation}
\frac{{\rm HD}(t)}{{\rm LD}} = \beta \left[1 - \gamma \frac{\int{L_{\rm FIR}(t)
{\rm d}t}}{L_{\rm FIR,int}}\right]~,
\label{eq:2}
\end{equation}
where $\gamma$ is the fraction of the initial HD component that is
consumed during the whole outburst, and $L_{\rm FIR,int}$ the FIR
luminosity integrated over the whole course of the outburst.  The
large-scale low-density component LD is assumed to remain unchanged.


The results of these simulations have been presented on the right
hand side in Fig.~\ref{fig:SB-evo}. The top panel shows the FIR
light curve of the outburst for three peak luminosities.  The
middle panel shows the luminosity of a representative high-density
component for the three \irlum \,curves. The bottom panel shows
the high-low-density ratio for these same FIR light curves.
Combining the results of this simulation with the data shows that
the HD/CO(1-0) data points in the left panels are a measure of the
evolution of the starburst. This implies that the OH MM sources
are galaxies in an early stage of star formation, which is
consistent with OH MM sources being found in starburst-dominated
galaxies \citep{GenzelEA1998, BaanK2006}.

\section{Chemistry}
\label{sec:chemistry} The model presented in the previous section
does well in explaining the evolution of the different gas
components, but it makes no distinction between the different
high-density tracers. Even though the emission of the different
molecules originates in the same regions, the line strengths are
influenced by the environmental properties like the (column)
density, temperature, and the type and strength of the prevailing
radiation field. In order to study the effects of these parameters
on the emission characteristics of the sources, we remove the
intrinsic difference in line strength between the galaxies in our
sample and use line ratios to find diagnostic properties.
Fig.~\ref{fig:3luik} presents the ratios of the integrated lines
of \HCOP/HCN, HNC/\HCOP\, and HNC/HCN against each other.

In order to interpret the behavior of the sources in this diagram, we
compare the data to the theoretical models that treat the chemistry
and radiative transfer of molecular clouds, including all the relevant
heating, cooling and chemical processes \citep[see][and Spaans \etal\
in these proceedings]{meijSpaans05,meijSI07}. A large grid of models
was created by \cite{meijSI07}, varying the strength of the radiation
field, its type (UV and X-ray), the gas density and the column
density. These results are compared to our observational data in
Fig.~\ref{fig:3luik} (note: not all models are shown, some fall out of
the range of our figure).

\subsection{PDR models}
\label{sec:pdr-models}
\begin{figure}
\begin{center}
 \includegraphics[width=11cm]{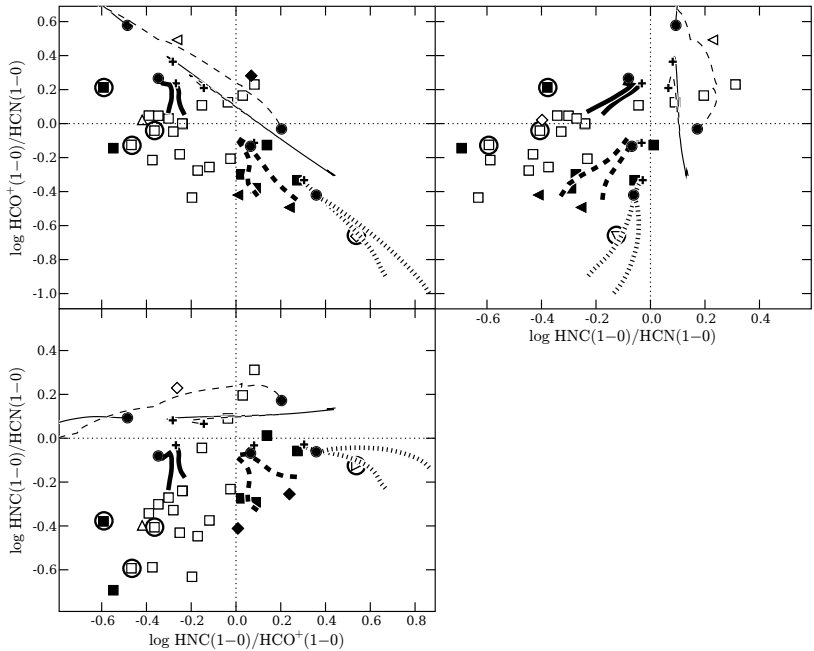}
  \caption{\textit{top left} Integrated \HCOP(1-0)/HCN(1-0) versus
HNC(1-0)/\HCOP(1-0) ratios. \textit{top right} Integrated
\HCOP(1-0)/HCN(1-0) versus HNC(1-0)/HCN(1-0) ratios.
\textit{bottom left} Integrated HNC(1-0)/HCN(1-0) versus
HNC(1-0)/\HCOP(1-0) ratios. Explanation about the plot symbols and
line styles is provided in Sect.~\ref{sec:chemistry}.}
  \label{fig:3luik}
\end{center}
\end{figure}

The results of the PDR models (UV radiation field) are shown in
Fig.~\ref{fig:3luik} with heavy lines, where the line styles indicate
different gas densities (solid: $n$=$10^{4.5}$, dashed:
$n$=$10^{5.0}$, and striped: $n$=$10^{5.5}$ cm$^{-3}$). The tracks
vary as a function of column density, which ranges from $N$=$10^{22}$
cm$^{-2}$ (the column density below which the strength of the emission
lines decreases rapidly) to $N$=$10^{23}$, $N$=$10^{23.5}$ and
$N$=$10^{24}$ cm$^{-2}$ for the $n$=$10^{4.5}$, $n$=$10^{5.0}$, and
$n$=$10^{5.5}$ cm$^{-3}$ models, respectively (corresponding to a
cloud size of 1 pc; indicated by the symbol at one end of the
tracks). Two different radiation field strengths are shown: $F_{\rm
UV}$=$1.6$ erg cm$^{-2}$ s$^{-1}$, indicated by a plus symbol at the
highest column density point, and $F_{\rm UV}$=$160$ erg cm$^{-2}$
s$^{-1}$, indicated with a circle.

Two observations can be made, when comparing the data and the
models. First of all one can see that the models are separated
based on the density ($n$) in the \HCOP/HCN and \HCOP/HNC line
ratios. The HNC/HCN line ratio shows no differentiation. This can
be explained in terms of the critical density of the individual
molecules. \HCOP\ has a critical density of about
$3$$\times$$10^5$ cm$^{-3}$, whereas the critical density of HCN
and HNC is around $3$$\times$$10^6$ cm$^{-3}$. Therefore the
excitation of \HCOP\ will differ from HCN and HNC for different
densities. A second observation that can be made is that most of
the OH MM sources cluster together in an area traced by PDR models
that have a high-density ($n$$\ge$$10^{5.5}$ cm$^{-3}$), and a
high column density $N$$\ge$$10^{22}$ cm$^{-2}$. This points to
the classical OH MM model of IR (UV radiation reprocessed by the
surrounding gas and dust) pumped, low (and variable) gain
amplification \citep{baan89}.

\subsection{XDR models}
\label{sec:xdr-models}
The results of two XDR simulations are also shown in
Fig.~\ref{fig:3luik}, using thin lines. Again, the line styles
indicate different gas densities (solid: $n$=$10^{5.5}$, and dashed:
$n$=$10^{6.0}$ cm$^{-3}$); column densities range from $N$=$10^{22}$
to $N$=$10^{24}$-$10^{24.5}$; and the radiation field strengths are
$F_{\rm X}$=$1.6$ (plus), and $F_{\rm X}$=$160$ erg cm$^{-2}$ s$^{-1}$
(circle).

The XDR models are not as well differentiated as the PDR models.
Because X-ray photons penetrate the molecular cloud much easier,
the XDRs do not show the strong density dependency seen for the
PDRs, making the distinction between different XDR models very
difficult. The addition of higher transitions and other molecules
(e.g. CN, CO$^+$, HOC$^+$) will most likely break this degeneracy.
Another problem with trying to identify XDR sources is that they
are in general smaller that PDR sources and thus are more affected
by beam dilution, especially in single dish observations like
ours. This will affect the detection rate of XDR sources and could
make hybrid sources look like PDR sources, even if the XDRs are
energetically more important than the PDRs.

Despite these drawbacks, the XDR models do provide diagnostics, since
they clearly separate from the PDR models in their HNC/HCN line
ratio. The PDR models approach, but never cross the HNC/HCN=1 line and
the XDR models all have HNC/HCN$>$1. Therefore, the HNC/HCN line ratio
is an excellent way to determine whether a system's chemistry is
dominated by UV or X-ray photons.

\subsection{Terra Incognita}
\label{sec:terra-incognita} Not all our observational data in
Fig.~\ref{fig:3luik} is covered by the models. The few sources in
our sample with known \water\ MM activity (indicated by plot
symbols surrounded by circles) are also located in this area,
which is characterized by lower HNC and higher \HCOP\ line
strengths compared to HCN. The fact that this area is not covered
by the models suggests that  other processes influence the line
ratios, such as strong shocks. Shocks are not treated in the
models and can have profound effects on the chemistry in molecular
clouds as they may selectively destroy HNC \citep{schilkeEA} and
enhance \HCOP\ relative to HCN \citep{dickinsonEA80}.  A
simultaneous increase in \HCOP\ and a decrease in HNC shifts the
PDR models to the uncovered region, implying that the \water\ MM
sources in our sample are UV driven systems with strong shocks.
This would suggest that these \water\ MM sources are similar to
shock-induced Galactic \water\ maser spots (see other
contributions in these proceedings).

\section{Conclusions}
\label{sec:conclusions} Molecular line emissions of multiple
species (and transitions) provide excellent diagnostics for
understanding the status of the nuclear gas in extra-galactic
sources. The different molecules trace different (density)
components and the ratio of high and low-density tracer lines
follows the star formation activity in the system. Comparing
different high-density tracers tells a lot about physical
characteristics of the gas. The \HCOP/HCN and \HCOP/HNC line
ratios are good proxies for the density of the gas, due to the
different critical densities of the species. PDR and XDR sources
can be distinguished using the HNC/HCN line ratio: PDR sources all
have ratios lower than unity and XDRs have ratios larger than 1.
OH MM sources cluster in a particular part of the diagnostic
diagram, which is only traced by PDR models with densities higher
than $10^{5.5}$ cm$^{-3}$, confirming the classical OH MM model of
FIR (UV radiation reprocessed by the surrounding gas and dust)
pumped amplification with low but variable gains.



\end{document}